\documentclass[a4paper,11pt]{article}
\usepackage{pos}
\usepackage{url}
\usepackage{amsmath}
\usepackage{gensymb}
\usepackage[utf8]{inputenc}
\usepackage{wrapfig}

\DeclareMathOperator{\fourq}{{\mathcal{O}_{\Gamma \Gamma}}}

\title{BSM $B - \bar{B}$ mixing on JLQCD and RBC/UKQCD $N_f=2+1$ DWF ensembles}

\author[a,b]{Peter Boyle}
\author[a]{Luigi Del Debbio}
\author*[a]{Felix Erben}
\author[c,d]{Andreas J\"uttner}
\author[e,f]{Takashi Kaneko}
\author[a]{Michael Marshall}
\author[a]{Antonin Portelli}
\author[g]{J Tobias Tsang}
\author[h]{Oliver Witzel}

\affiliation[a]{Higgs Centre for Theoretical Physics, School of Physics \& Astronomy,The University of Edinburgh,
 Edinburgh EH9 3FD, United Kingdom}

\affiliation[b]{Physics Department, Brookhaven National Laboratory, Upton, NY, USA}

\affiliation[c]{Physics and Astronomy, University of Southampton, Southampton SO17 1BJ, UK}

\affiliation[d]{Theoretical Physics Department, CERN, 1211 Geneva 23, Switzerland}

\affiliation[e]{High Energy Accelerator Research Organization (KEK), Tsukuba 305-0801, Japan}

\affiliation[f]{School of High Energy Accelerator Science, SOKENDAI (The Graduate University for Advanced Studies), Ibaraki 305-0801, Japan}

\affiliation[g]{CP$^3$-Origins and IMADA, University of Southern Denmark, Campusvej 55,DK-
5230 Odense M, Denmark}

\affiliation[h]{Center for Particle Physics Siegen, Theoretische Physik 1, Naturwissenschaftlich-Technische Fakultät,Universität Siegen, 57068 Siegen, Germany}

\emailAdd{felix.erben@ed.ac.uk}
\emailAdd{tsang@imada.sdu.dk}

\abstract{We are presenting our ongoing Lattice QCD study on $B - \bar{B}$ mixing on several RBC/UKQCD and JLQCD ensembles with 2+1 dynamical-flavour domain-wall fermions, including physical-pion-mass ensembles. We are extracting bag parameters $B_{B_d}$ and $B_{B_s}$ using the full 5-mixing-operator basis to study both Standard-Model mixing as well as Beyond the Standard Model mixing, using a fully correlated combined fit to two-point functions and ratios of three-point and two-point functions. Using 15 different lattice ensembles we are simulating a range of heavy-quark masses from below the charm-quark mass to just below the bottom-quark mass.}

\FullConference{%
 The 38th International Symposium on Lattice Field Theory, LATTICE2021
  26th-30th July, 2021
  Zoom/Gather@Massachusetts Institute of Technology
}

\begin{document}
\maketitle

\section{Introduction}
\begin{figure}[htbp] 
  \centering
\includegraphics[width=0.34\textwidth]{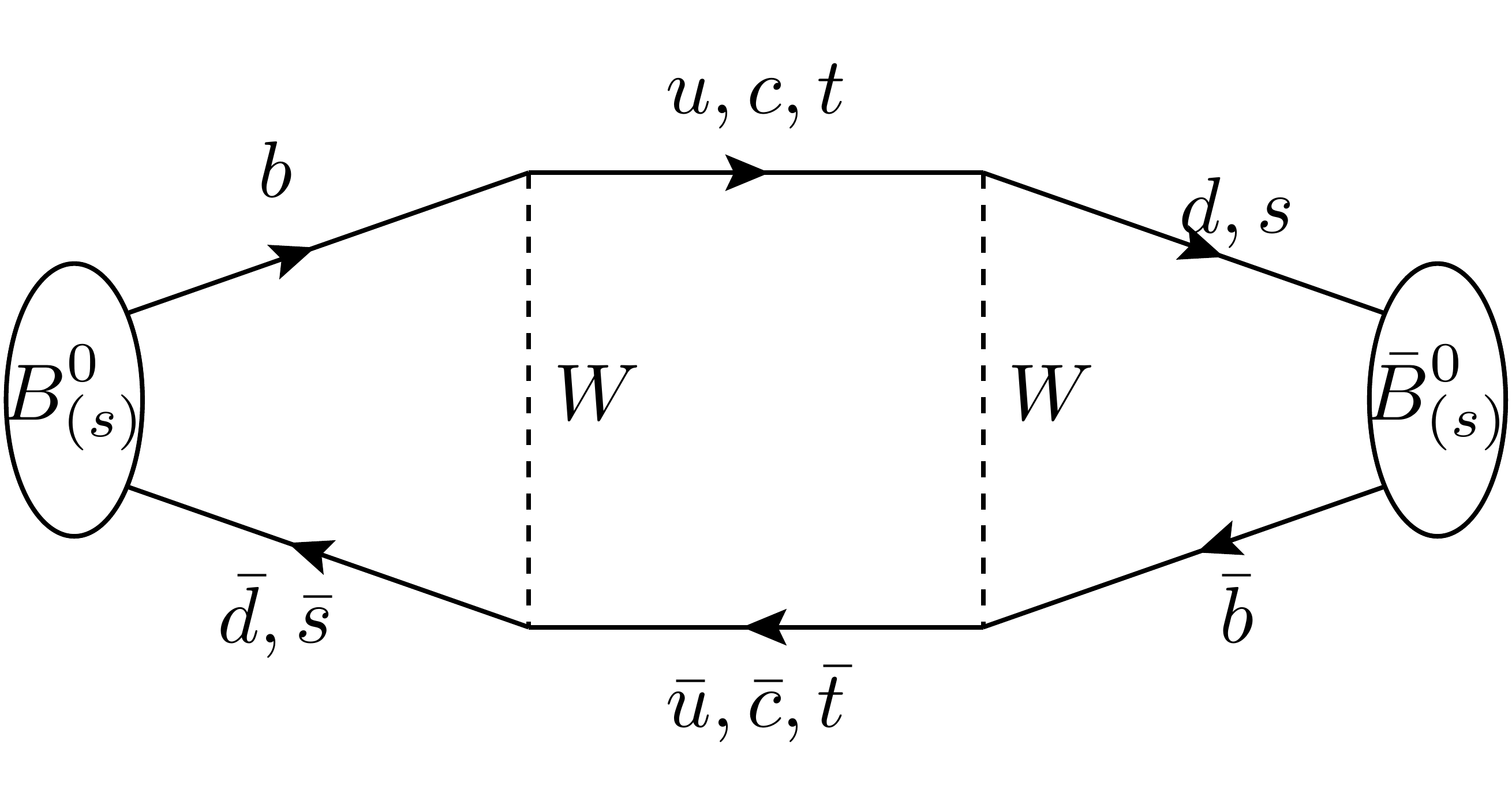}
\hspace{1cm}
\includegraphics[width=0.34\textwidth]{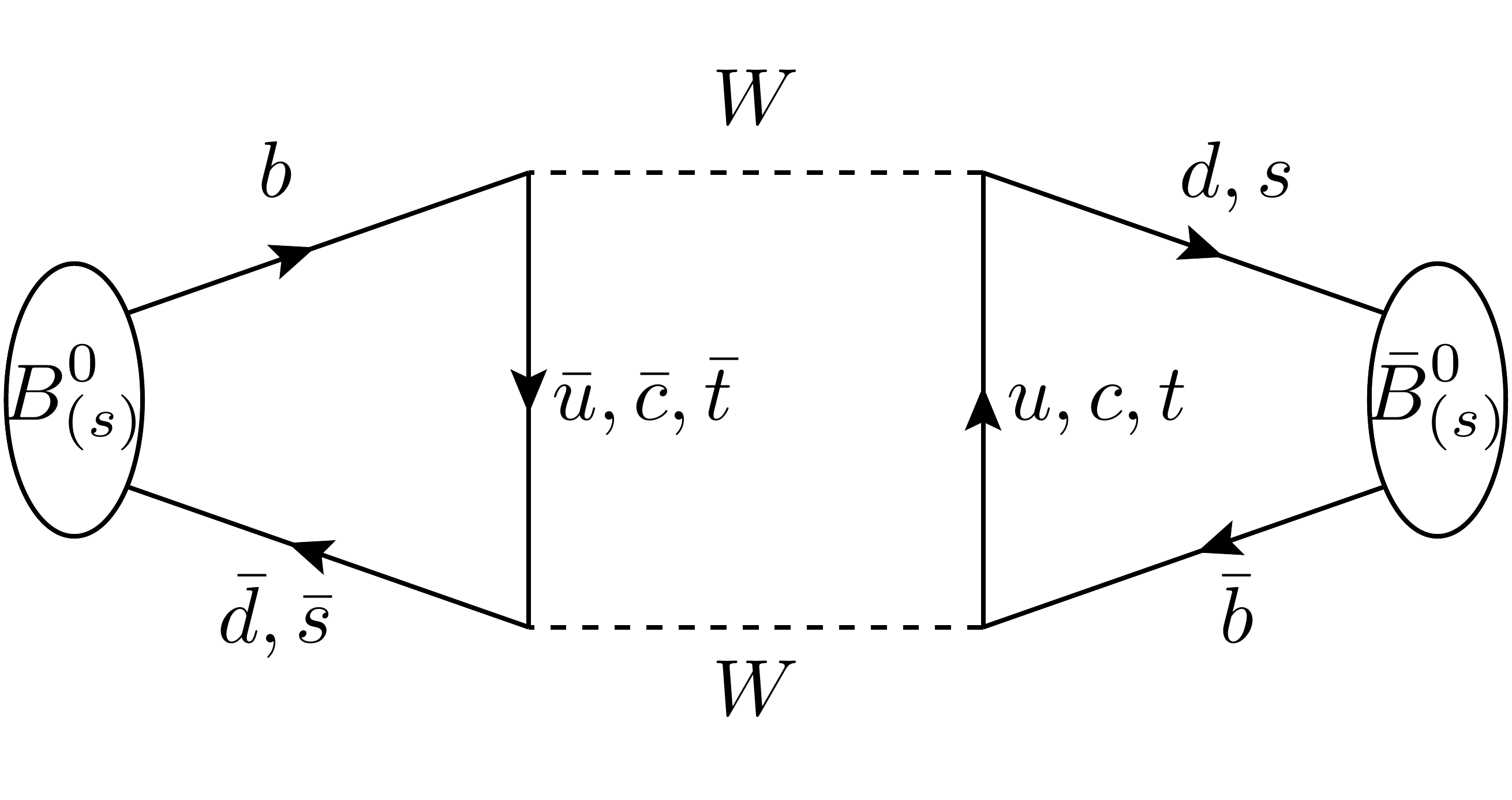}
  \caption{The two $W$-exchange box diagrams contributing to $B-\bar{B}$ mixing. Both are dominated by the $t$ quark loop.} 
  \label{figure:BB-mixing}
\end{figure}

Neutral $B_{(s)}$ meson mixing occurs at the one-loop level in the Standard Model (SM) via the box diagrams shown in Fig.~\ref{figure:BB-mixing}.
Contributions with a top quark dominate, rendering these processes inherently short-distance. Hence lattice QCD calculations are well suited to determine the non-perturbative contributions due to the strong force, and meson mixing can be expressed in terms of local four-quark operators with  $\Delta b=2$. The calculation is similar to the case of $K-\bar{K}$ mixing, which the RBC/UKQCD collaboration has studied extensively \cite{Garron:2016mva,Kettle:2016mmt,Boyle:2017skn,Boyle:2017ssm,Boyle:2018eor}. $B_{(s)}-\bar{B}_{(s)}$ mixing, however, faces the additional challenge to simulate the much heavier $b$-quarks. In the past we considered $b$ quarks in the static limit \cite{Albertus:2010nm,Aoki:2014nga}, but more recently moved to a fully relativistic setup \cite{Boyle:2018knm}. Building upon \cite{Boyle:2018knm} we report here on recent extensions: 1) We consider the full basis with five operators needed in some extensions of the SM. 2) We are working towards a fully non-perturbative renormalisation (NPR) \cite{Martinelli:1994ty} using the RI-SMOM scheme \cite{Sturm:2009kb}. 3) Measurements on additional ensembles are included extending the range of heavy-quark masses simulated and giving us a better handle to estimate systematic uncertainties.

This will enable us to determine from first principles several quantities which provide stringent tests of the SM or constrain Beyond the Standard Model (BSM) physics. A simple example is the comparison of experimental and theoretical determinations of the mass differences of the neutral $B_{(s)}$ mesons, $\Delta M_d$ and $\Delta M_s$. HFLAV \cite{HFLAV:2019otj} provides the precise average of the experimental results \cite{ARGUS:1987xtv,CDF:2006imy,LHCb:2011vae,LHCb:2012dgy,LHCb:2012mhu,LHCb:2013lrq,LHCb:2013fep,LHCb:2014iah} and on  the lattice HPQCD \cite{Dowdall:2019bea} and Fermilab/MILC \cite{FermilabLattice:2016ipl} have determined   $\Delta M_d$ and $\Delta M_s$. Further determinations based on QCD sum rules \cite{Grozin:2016uqy,Grozin:2017uto,Grozin:2018wtg,Kirk:2017juj,King:2019lal} exist. Currently this comparison \cite{DiLuzio:2019jyq} shows a tension between the  lattice results. While HPQCD is in agreement with the experimental value \cite{HFLAV:2019otj}, Fermilab/MILC is not. Our own work \cite{Boyle:2018knm} provides so far only the ratio $\Delta M_d / \Delta M_s$, where renormalisation coefficients cancel.

Our work is based on  $N_f=2+1$ domain-wall fermion (DWF) \cite{Kaplan:1992bt,Blum:1996jf,Blum:1997mz} gauge field ensembles generated by  the RBC/UKQCD \cite{Allton:2008pn,PhysRevD.93.074505,Boyle:2017jwu} and JLQCD \cite{Nakayama:2016atf} collaborations. Some of their properties are listed in Table \ref{table:ensembles}.
\begin{table}
\centering
\begin{tabular}{|c|c|c|c|c|c|c|c|c|}
\hline 
 & $L/a$ & $T/a$ & $a^{-1}$ [GeV] & $m_\pi$ [MeV] & $m_\pi L$ & hits $\times N_\mathrm{conf}$ & collaboration id\\ 
\hline a1.7m140 & 48 & 96 & 1.730(4) & 139.2 & 3.9 & $48 \times 90$ & R/U C0\\ 
a1.8m340 & 24 & 64 & 1.785(5) & 339.8 & 4.6 & $32 \times 100$& R/U C1\\
a1.8m430 & 24 & 64 & 1.785(5) & 430.6 & 5.8 & $32 \times 101$& R/U C2\\ 
\hline 
a2.4m140  & 64 & 128 & 2.359(7) & 139.3 & 3.8 & $64 \times 82$& R/U M0\\ 
a2.4m300  & 32 & 64 & 2.383(9) & 303.6 & 4.1 & $32 \times 83$& R/U M1\\ 
a2.4m360 & 32 & 64 & 2.383(9) & 360.7 & 4.8 & $32 \times 76$& R/U M2\\
a2.4m410  & 32 & 64 & 2.383(9) & 411.8 & 5.5 & $32 \times 81$& R/U M3\\ 
\hline 
 a2.5m230-L & 48 & 96 & 2.453(4) & 225.8 & 4.4 & $24 \times 100$ & J C-ud2-sa-L\\ 
 a2.5m230-S & 32 & 64 & 2.453(4) & 229.7 & 3.0 & $16 \times 100$ & J C-ud2-sa\\ 
 a2.5m310-a & 32 & 64 & 2.453(4) & 309.1 & 4.0 & $16 \times 100$ & J C-ud3-sa\\ 
a2.5m310-b & 32 & 64 & 2.453(4) & 309.7 & 4.0 & $16 \times 100$ & J C-ud3-sb\\ 
\hline 
a2.7m230 & 48 & 96 & 2.708(10) & 232.0 & 4.1 & $48 \times 72$ & R/U F1M\\ 
\hline 
a3.6m300-a & 48 & 96 & 3.610(9) & 299.9 & 3.9 & $24 \times 50$ & J M-ud3-sa\\ 
a3.6m300-b & 48 & 96 & 3.610(9) & 296.2 & 3.9 & $24 \times 50$ & J M-ud3-sb\\ 
\hline 
a4.5m280 & 64 & 128 & 4.496(9) & 284.3 & 4.0 & $32 \times 50$ & J F-ud3-sa\\ 
\hline 
\end{tabular} 
\caption{List of ensembles used in this work. Both the RBC/UKQCD and the JLQCD ensembles feature three lattice spacings which together range from $a^{-1}=1.7$ GeV down to $a^{-1}=4.5$ GeV. To have a consistent naming convention in our set of ensembles from two collaborations, we introduce a special shorthand notation in the first column which is used throughout this work. For readers familiar with earlier work by the RBC/UKQCD ("R/U") and JLQCD ("J") collaborations, we also list in the last column names previously used by the respective collaborations. The column `hits $\times N_\text{conf}$' refers to the number of $Z_2$ wall sources placed equidistantly on each configuration.}
  \label{table:ensembles}
\end{table}
These ensembles feature pion masses from $m_\pi=430$ MeV down to the physical range of $m_\pi=139$ MeV and six values of the lattice spacing ranging from $a^{-1}=1.7$ GeV up to $a^{-1}=4.5$ GeV. In addition to the two ensembles at a physical pion mass, there is one dedicated pair to study finite volume effects (all parameters the same but the box size is reduced from $m_\pi L =4.4$ down to $m_\pi L =3.0$) and two other pairs bracketing the strange quark mass to investigate the effect of the strange sea-quark mass. On all ensembles we simulate multiple heavy-quark masses from below or around $m_c$ up to just below $m_b$ on the finest JLQCD ensemble with $a^{-1}=4.5$ GeV. By performing a combined analysis in terms of a global fit, we expect excellent control when taking the continuum limit and extrapolating to physical quark masses. 

Light and strange quarks are simulated using the same DWF action as was used in the sea sector. Heavy quarks ranging from charm to bottom are simulated using stout-smeared \cite{Morningstar:2003gk} M\"obius DWF \cite{Brower:2012vk} with parameters $b=1.5$ and $c=0.5$. All our computations are done using the software suites Grid \cite{Boyle:2016lbp} and Hadrons \cite{Portelli:Hadrons}.

\section{Lattice computation}
One ingredient we compute on the lattice are two-point correlation functions of mesons with a light and a heavy quark, and mesons with a strange and a heavy quark. These are given by
\begin{align}
C_{\Gamma_1,\Gamma_2}^{s_1,s_2} = \sum_{\mathbf{x}} \langle O_{\Gamma_2}^{s_2}(\mathbf{x},t) O_{\Gamma_1}^{s_1}(\mathbf{0},0)^\dagger \rangle = \sum_{n=0}^\infty \frac{M_{\Gamma_2 n}^{s_2} M_{\Gamma_1 n}^{s_1 *}}{2 E_n} (e^{-E_n t} \pm e^{-E_n (T-t)})
\, ,
\end{align}
with the energy $E_n$ and matrix element $M_{\Gamma_i n}^{s_i}=\langle X_n | O_{\Gamma_i}^{s_i} | 0 \rangle$ of the $n^\mathrm{th}$ excited meson state $X_n$. The $\pm$ sign depends on the choice of interpolation operators
\begin{align}
O_{\Gamma_i}^{s_i} (\mathbf{x},t) = \bar{q}_2(\mathbf{x},t) \sum_{\mathbf{y}} \omega_s(\mathbf{x},\mathbf{y}) \Gamma_i q_1(\mathbf{y},t)
\, ,
\end{align}
which are defined by their quark content $q_1,q_2$ and their Dirac structure, which we limit to $\Gamma_i=\gamma_5 \equiv P$ (pseudoscalar) and $\Gamma_i=\gamma_0 \gamma_5 \equiv A$ (temporal component of the axial vector). The smearing operator $\omega_s$ is chosen to be either smeared ($S$) or local ($L$) at source and sink of each propagator. We use Gaussian smearing \cite{GUSKEN1989266,ALEXANDROU199160,UKQCD:1993gym} on the coarser a1.7 to a2.4 ensembles, where we have correlation functions $C^{s_1,s_2}$ with $s_1,s_2 \in \{ SL, SS \}$ for the light and strange propagators (the first entry corresponds to the source, the second to the sink) and $s_1,s_2 = LL$ for the heavy-quark propagators. On the finer a2.5 to a4.5 ensembles, we only have two-point functions with local interpolators at both source and sink ($LL$).

\begin{figure}[tbp] 
  \centering
  \includegraphics[width=0.34\textwidth]{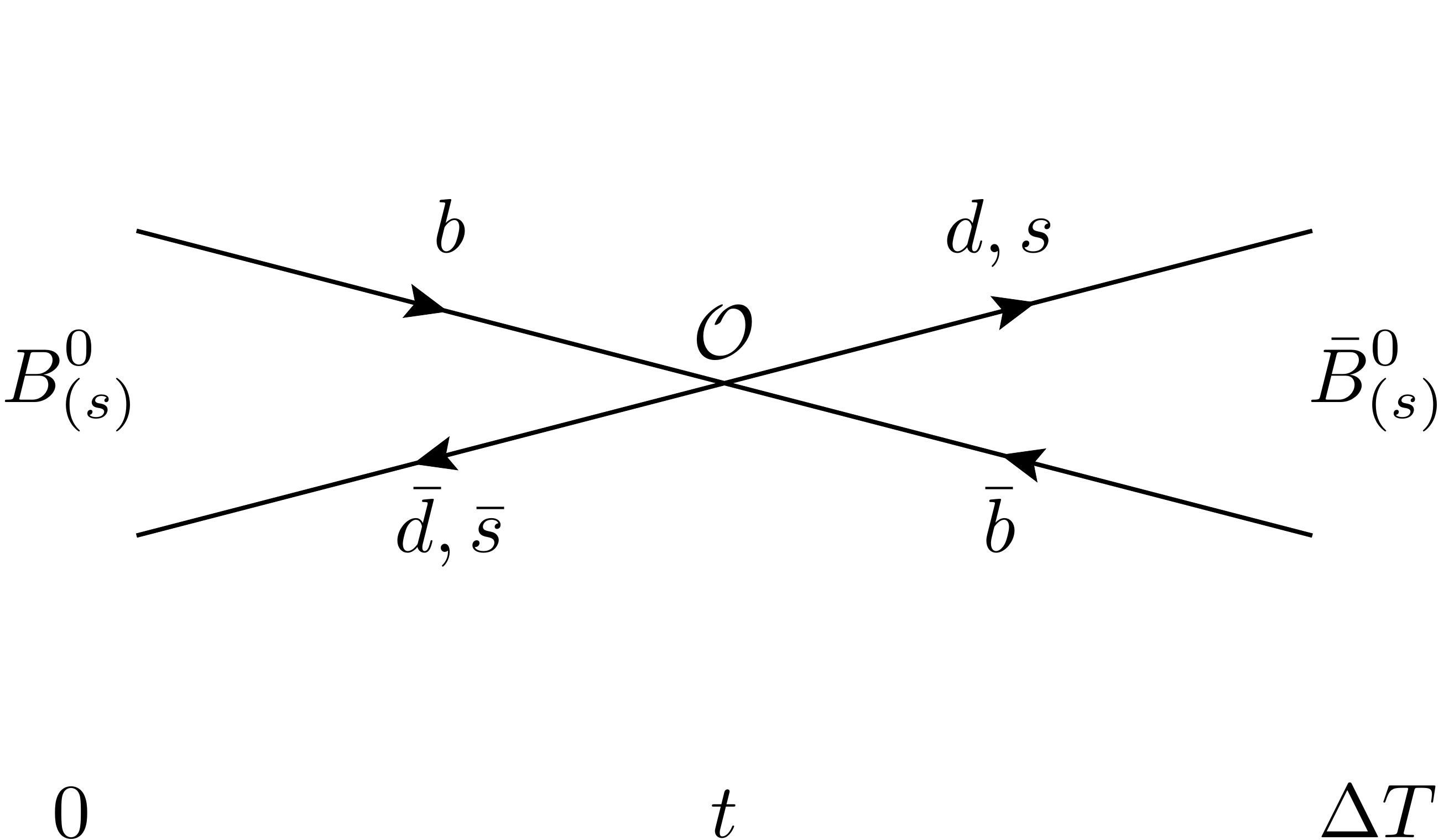}
  \caption{Sketch of the lattice setup to calculate three-point functions with a four-quark-operator insertion at time $t$. The time separation between the $B_{(s)}^0$ and $\bar{B}_{(s)}^0$ mesons is $\Delta T$.}  
  \label{figure:BB-lattice}
\end{figure}

We determine the non-perturbative contributions to neutral $B_{(s)}$ meson mixing by implementing the four-quark operators
\begin{align}
\fourq=(\bar{b}_a \Gamma d_a) (\bar{b}_b \Gamma d_b)
\, ,
\end{align}
and calculate on the lattice three-point correlation functions as schematically shown in Fig.~\ref{figure:BB-lattice}.
These three-point functions are described by
\begin{align}
\nonumber C_3^{\fourq}(t,\Delta T) &=  \langle P(\Delta T) {\fourq}(t) \bar{P}^\dagger(0) \rangle \\
\nonumber & \approx \frac{P_0^2}{4E_0^2} \langle X_0 | {\fourq} | X_0 \rangle e^{-E_0 \Delta T}\times  \\
 & \bigg[ 1+  2 \frac{P_1 E_0}{P_0 E_1} \frac{\langle X_0 | {\fourq} | X_1 \rangle}{\langle X_0 | {\fourq} | X_0 \rangle} e^{-\Delta E \Delta T /2}  \cosh \big[ \Delta E \big( t- \Delta T /2\big) \big] \bigg] 
 \, ,
 \label{eqn:3pt}
\end{align}
with $\Delta E = E_1-E_0$. In Eq.~(\ref{eqn:3pt}) we truncate the expressions by writing only the ground and the first excited states explicitly. Further we use the shorthand $P_n=M_{\Gamma_5 n}^{s}$ with $s=S$ on the a1.7 to a2.4 ensembles and $s=L$ on the a2.5 to a4.5 ensembles. The mixing operators $\fourq$ are
\begin{align*}
\mathcal{O}_1 =\mathcal{O}_{VV+AA} \, , \
\mathcal{O}_2 =\mathcal{O}_{VV-AA} \, , \
\mathcal{O}_3 =\mathcal{O}_{SS-PP} \, , \
\mathcal{O}_4 =\mathcal{O}_{SS+PP} \, , \
\mathcal{O}_5 =\mathcal{O}_{TT} \, ,
\end{align*}
where $\mathcal{O}_1$ is SM operator and $\mathcal{O}_{2-5}$ are important in several SM extensions \cite{Boyle:2017skn}. A great advantage of the DWF action is that due to the chiral symmetry the mixing between those operators is minimised: $\mathcal{O}_1$ does not mix with the others,  $\mathcal{O}_2$ mixes only with  $\mathcal{O}_3$ and  $\mathcal{O}_4$ mixes only with  $\mathcal{O}_5$.

\section{Fitting strategy}
\begin{figure}[tbp] 
  \centering
\includegraphics[width=0.99\textwidth]{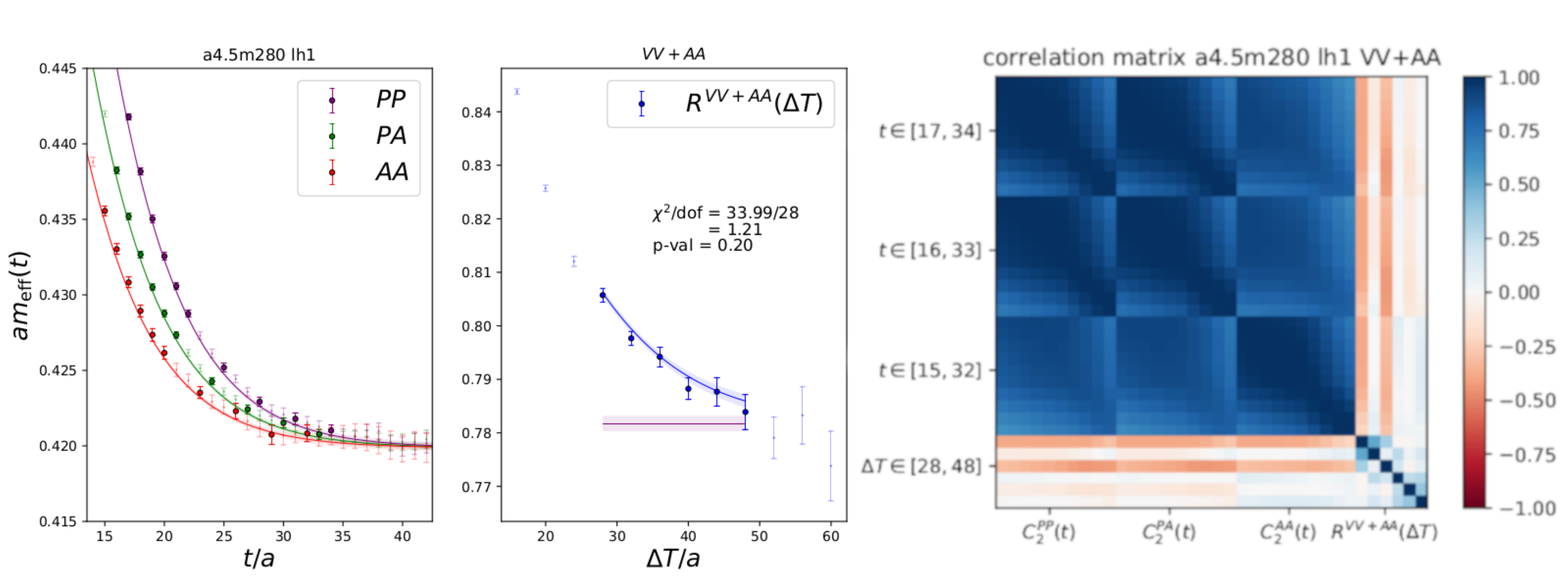}
  \caption{Data and reconstructed fit function of the lightest heavy-light meson on the a4.5m280 ensemble for the SM mixing operator $\mathcal{O}_1$. The left panel shows the three two-point functions, the middle panel shows the ratio of three-point and two-point functions, with the purple fit band showing the bag parameter obtained from the fit. Data points shown in bold are the ones actually used in the fit. In the two point function, all data points between $t^\mathrm{min}$ and a transition time slice $t^\mathrm{inter}$ are taken into account, above that only every $3\mathrm{rd}$ up to $t^\mathrm{max}$. Only one $\chi^2/$dof value is quoted as this is a single, correlated, combined fit to all data shown on both panels. The right panel shows the corresponding correlation matrix, where the large blue square on the top left belongs to the three two-point functions, which nicely decouple their correlations from the ratios.}
  \label{figure:fit-F1-lh1}
\end{figure}

\begin{figure}[htbp] 
  \centering
\includegraphics[width=0.79\textwidth]{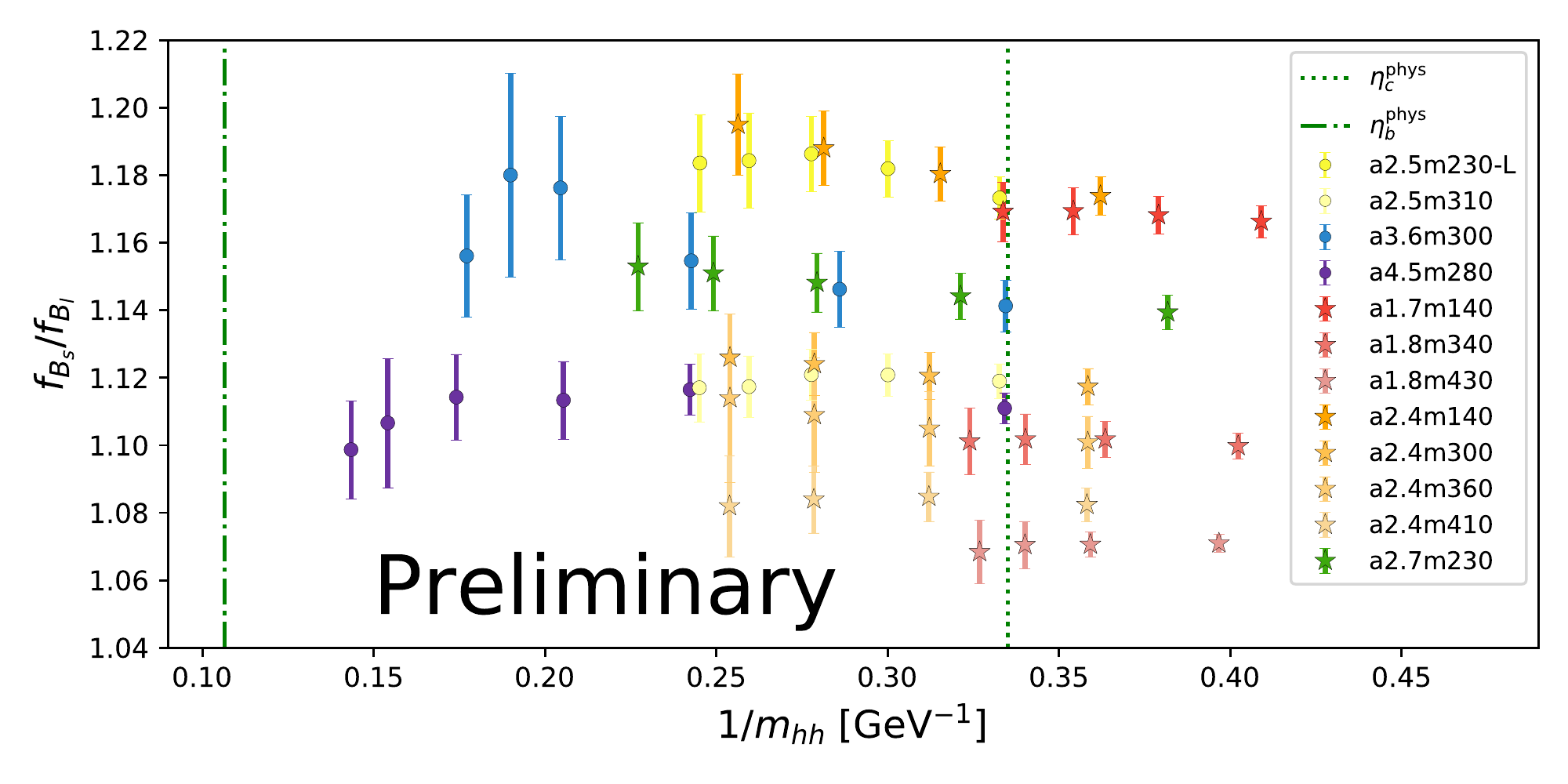}
\includegraphics[width=0.79\textwidth]{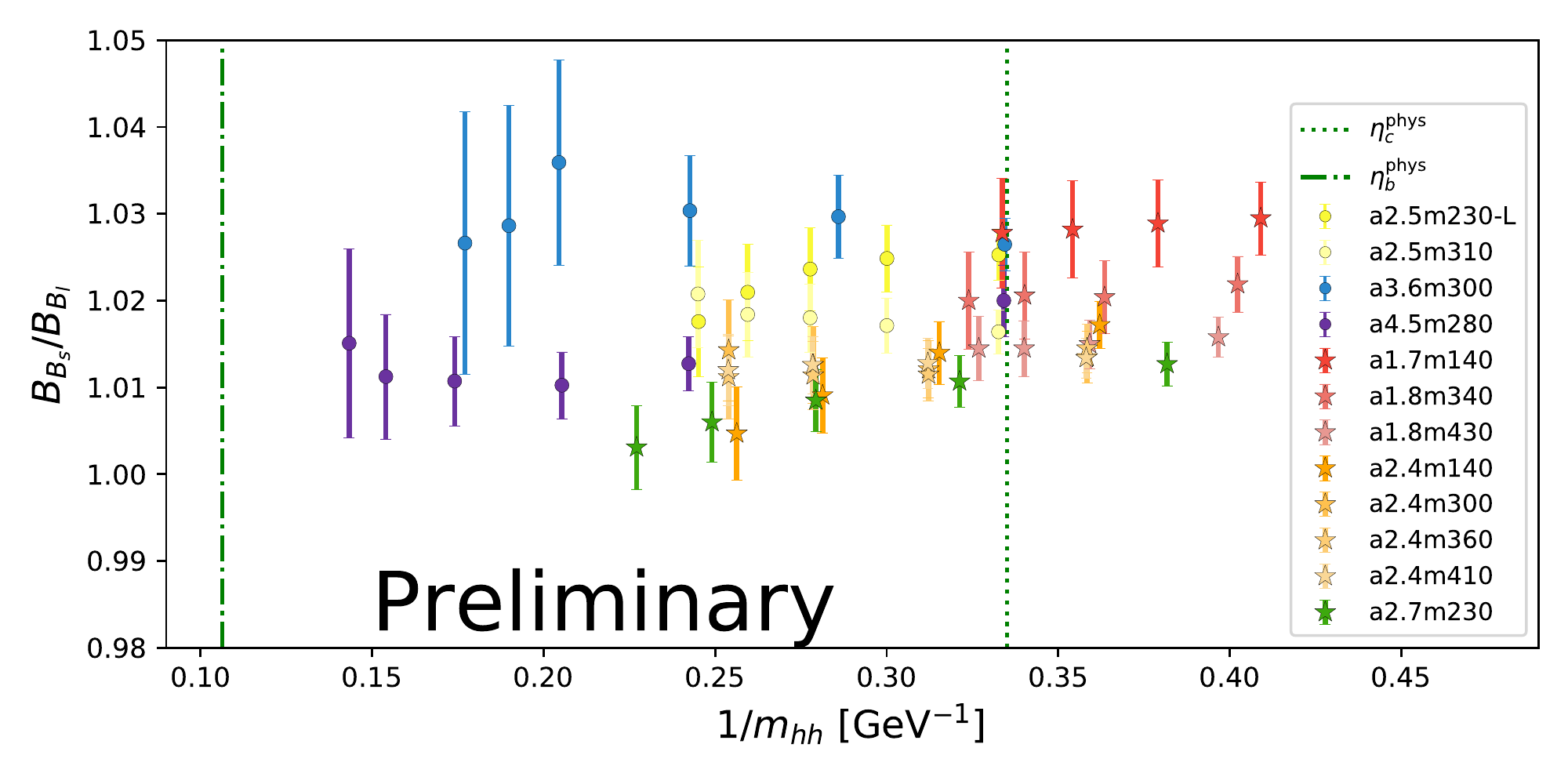}
  \caption{Ratios of decay constants $f_{B_s}/f_{B_l}$ in the top panel and ratio of bag parameters $B_{B_s}/B_{B_l}$ in the bottom panel, both for the SM operator $\mathcal{O}_1$, plotted against the inverse heavy-heavy pseudoscalar mass as a proxy for the heavy-quark mass. The left vertical green line is at the physical $\eta_b$ and the right vertical line at the physical $\eta_c$. Only the JLQCD data (a2.5, a3.6, a4.5) is fitted with the combined fit described in the text. The RBC/UKQCD data (a1.7, a1.8, a2.4, a2.7) is the data from \cite{Boyle:2018knm}, which used a different fit method. Using the method described here also for all ensembles is currently being worked on. In the $SU(3)$ symmetric limit ($m_\pi \to m_K$) these ratios are expected to be $1$, and ensembles closer to the physical $m_\pi$ can be seen to have ratios further away from $1$. At least in the $f_{B_s}/f_{B_l}$ ratio, discretisation effects are small, as can be seen by comparing the red a1.7m140 and orange a2.4m140 data.}
  \label{figure:ratios}
\end{figure}

\begin{figure}[htbp] 
  \centering
\includegraphics[width=0.49\textwidth]{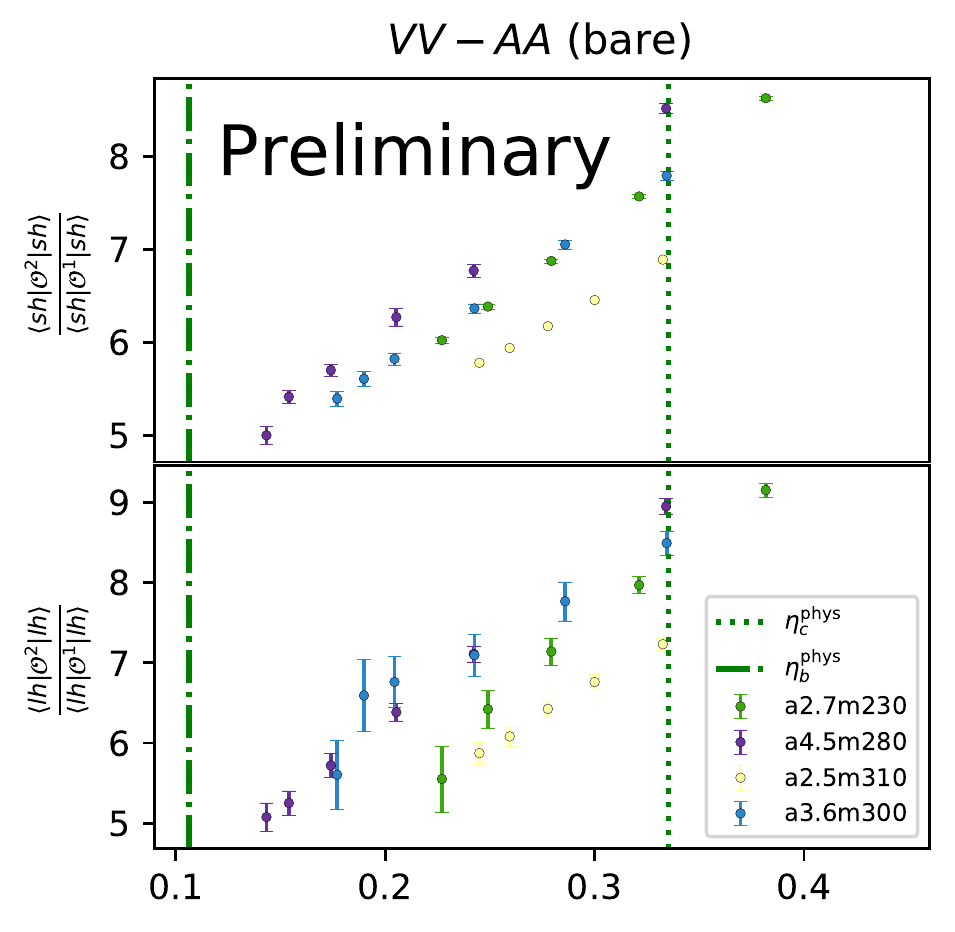}
\includegraphics[width=0.49\textwidth]{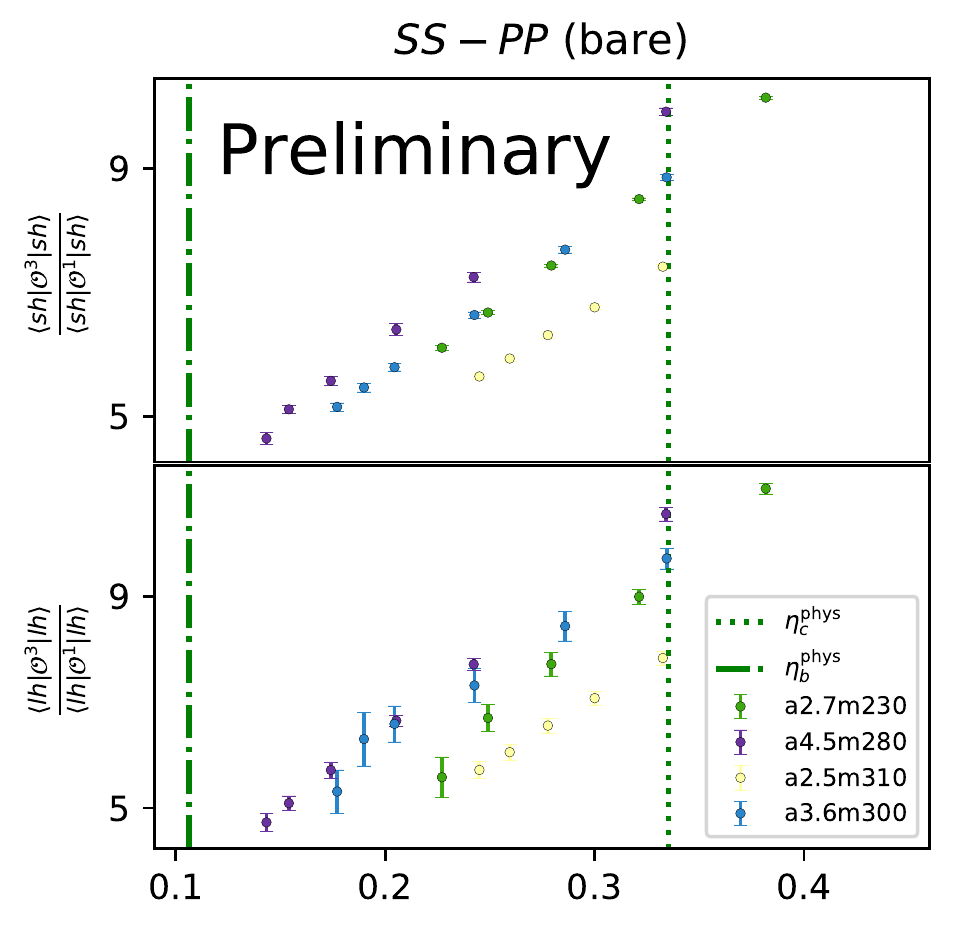}
\includegraphics[width=0.49\textwidth]{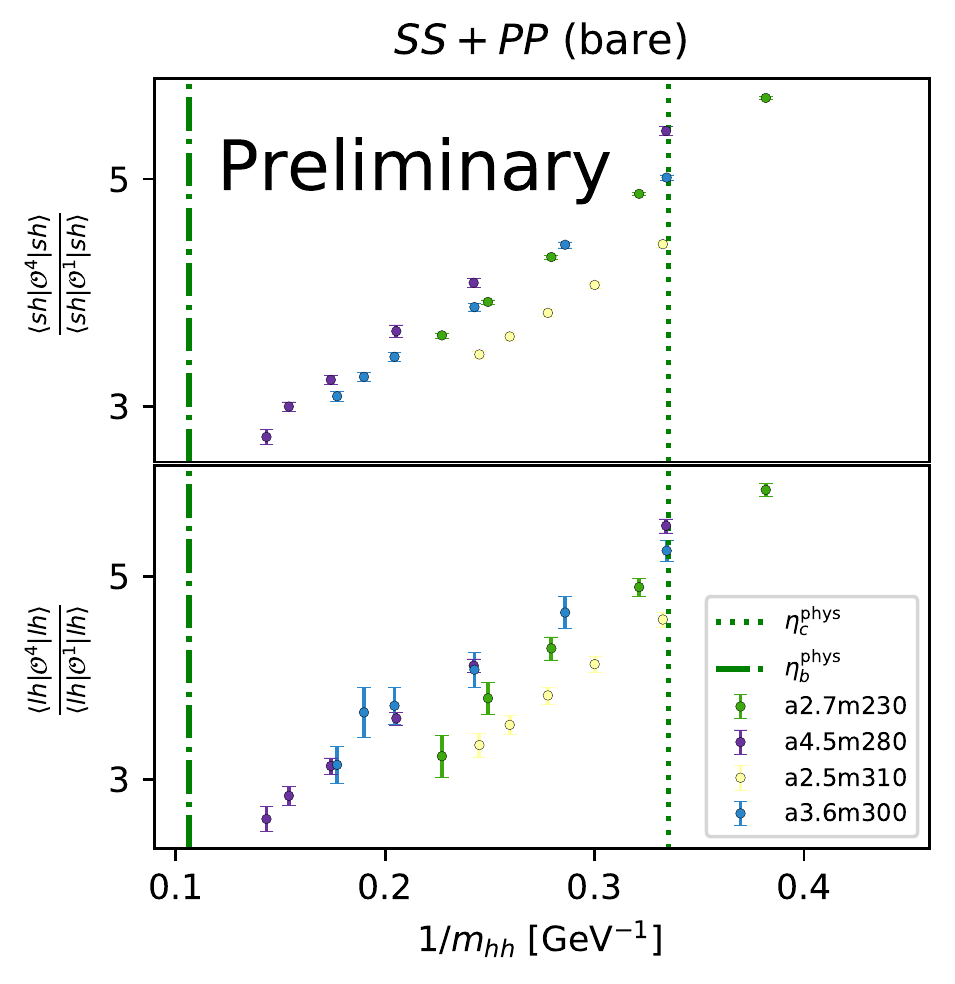}
\includegraphics[width=0.49\textwidth]{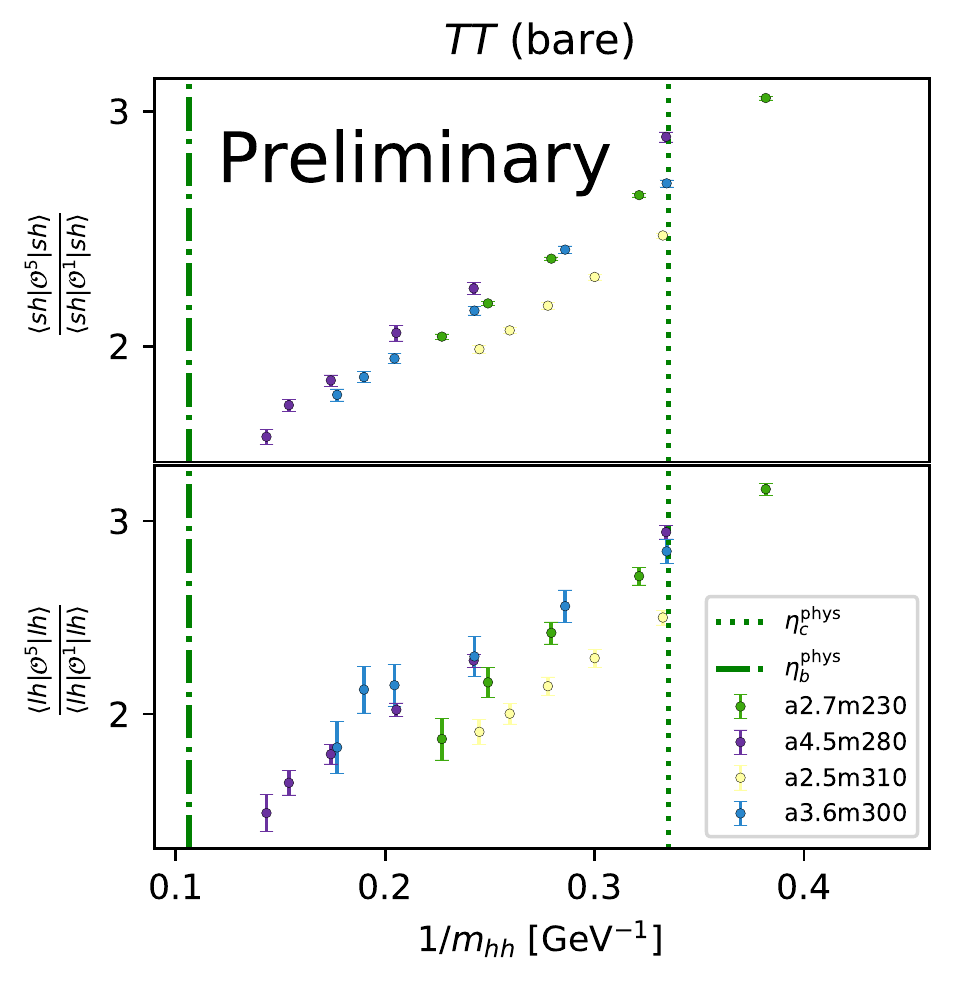}
\caption{Ratio of matrix elements with BSM operators $O_2$--$O_5$ over the matrix element with the SM operator $O_1$. The top panels show the ratios for strange-heavy mesons, and the bottom panels show the ratios for light-heavy mesons.}
  \label{figure:BSMratios}
\end{figure}

We are interested in the bag parameters
\begin{align}
B_i=\frac{\langle P | \mathcal{O}_i | \bar{P}^\dagger \rangle }{\langle P |  \mathcal{O}_i | \bar{P}^\dagger \rangle_{VSA}}
\, ,
\end{align}
which are defined as the ratio of a three-point-function matrix element over its vacuum-saturation approximation (VSA). At leading order the SM bag parameter is given by
\begin{align}
B_1=\frac{\langle P| \mathcal{O}_1 |\bar{P}^\dagger \rangle }{8/3 m_P^2f_P^2}
\, ,
\end{align}
with meson mass $m_P$ and decay constant $f_P$. The other bag parameters are 
\begin{align}
B_i=\frac{(m_q^2 + m_h^2)\langle P| \mathcal{O}_i | \bar{P}^\dagger \rangle }{N_i m_P^4f_P^2}
\end{align}
with quark masses $m_q \in \{ m_l, m_s \}, m_h$ and the normalisation factors $N_2=-5/3, N_3=1/3, N_4=2, N_5=2/3$ \cite{Boyle:2017skn}. The product of two two-point functions
\begin{align}
\nonumber C_{\Gamma_1,\Gamma_2} (t) C_{\Gamma_3,\Gamma_4}(\Delta T -t) &\approx \frac{M_{\Gamma_2 0} M_{\Gamma_1 0} M_{\Gamma_4 0} M_{\Gamma_3 0}}{4E_0^2}  e^{-E_0 \Delta T}\times  \\
 & \bigg[ 1+   \frac{E_0}{E_1}\bigg(\frac{M_{\Gamma_2 1} M_{\Gamma_1 1}}{M_{\Gamma_2 0} M_{\Gamma_1 0}} + \frac{M_{\Gamma_4 1} M_{\Gamma_3 1}}{M_{\Gamma_4 0} M_{\Gamma_3 0}} \bigg) e^{-\Delta E \Delta T /2}  \cosh \big[ \Delta E \big( t- \Delta T /2\big) \big] \bigg] 
 \, ,
 \label{eqn:2pt-prod}
\end{align}
has a very similar time behaviour to the three-point functions defined in Eq. \eqref{eqn:3pt}. We have omitted the smearing index in the matrix elements $M_{\Gamma i}^{s_i}$, which are chosen to cancel the overlap factors of the corresponding three-point function. This leads to the definition of ratios 
\begin{align}
R_1(t,\Delta T) & = \frac{C_3^{\mathcal{O}_1}(t,\Delta T)}{8/3 C_{PA} (t) C_{AP}(\Delta T -t)}\, , \\
R_i(t,\Delta T) & = \frac{C_3^{\mathcal{O}_i}(t,\Delta T)}{N_i C_{PP} (t) C_{PP}(\Delta T -t)} 
\, , \phantom{aa} 2 \leq i \leq 5
\, ,
\end{align}
which have a very good overlap with the bag parameters. In both Eq. \eqref{eqn:3pt} and Eq. \eqref{eqn:2pt-prod}, the $\cosh$-term is equal to $1$ for $t=\Delta T/2$, so that we can define $R_i(\Delta T) \equiv R_i(t=\Delta T/2,\Delta T)$. We have explored several strategies to extract the bag parameters from the two-point and three-point functions and have settled on a single, fully correlated, combined fit to
\begin{align*}
C_{PP}^{LL},C_{PA}^{LL},C_{AA}^{LL},R_i(\Delta T)
\end{align*}
for each individual bag parameter on the a2.5 to a4.5 ensembles. We have not performed the combined fits to the a1.7 to a2.4 ensembles yet, but we are planning to perform a similar combined fit to
\begin{align*}
C_{PP}^{SS},C_{PA}^{SS},C_{AA}^{SS},C_{PP}^{SL},C_{PA}^{SL},C_{AA}^{SL},R_i(\Delta T)
\end{align*}
on those. To achieve a fully correlated fit in the two-point functions, we are thinning out the correlation function above a transition value $t^\mathrm{inter}$ and use only every $3\mathrm{rd}$ timeslice up to the end of the fitrange $t^\mathrm{max}$. From the minimum fit timeslice $t^\mathrm{min}$ up to $t^\mathrm{inter}$, all timeslices are used. We observe that thinning leads to a better conditioned covariance matrix and more stable correlated fits. For the ratios $R_i(\Delta T)$ we fit all the data we have available in a range $[\Delta T^\mathrm{min},\Delta T^\mathrm{max}]$, which effectively corresponds to a thinning in the $\Delta T$ direction, as we have measured the three-point functions only for a subset of $\Delta T$ (typically every $4^\mathrm{th}$ value, but on a2.7m230 on every $2^\mathrm{nd}$). An example fit of the lightest heavy-light meson on the a4.5m280 for the SM bag parameter ensemble is shown in Fig. \ref{figure:fit-F1-lh1}. The figure also shows the correlation matrix corresponding to the fit, which shows that the ratios and the two-point functions are nicely decorrelated, making this combined fit possible. Earlier attempts to fit the two-point functions directly combined with the raw three-point functions were unstable due to too strong correlations between them.
We do a fit like this on each ensemble and for every meson (heavy-light and heavy-strange for $4-6$ different heavy masses on each ensemble) and for all five mixing operators.  In all cases we can choose fit ranges yielding a good correlated $\chi^2/$dof. The RBC/UKQCD ensembles (a1.7, a1.8, a2.4, a2.7) are tuned to be at the physical strange-quark mass. On the JLQCD ensembles, we have two pairs of ensembles (a2.5m310-a/b and a3.6m300-a/b) which differ only in their strange-quark mass, and we interpolate to the physical value using $2m_K^2-m_\pi^2$. We find that the effect of the different $m_s$ in those ensembles is mild. We show the fit results for the ratio of decay constants $f_{B_s}/f_{B_l}$ and the ratio of bag parameters $B_{B_s}/B_{B_l}$ in Fig.~\ref{figure:ratios}.
This figure illustrates the far reach in the heavy-quark mass possible within our setup, which is facilitated, in particular, through the inclusion of the JLQCD ensembles with very fine lattice spacings. In both ratios, the heavy-quark mass dependence is very mild for every ensemble. We quote the ratios and not the individual decay constants or bag parameters as the NPR is not finalised yet. We also show the bare matrix elements $\langle P | \mathcal{O}_i | \bar{P} \rangle$ for the BSM operators in Fig. \ref{figure:BSMratios}.

\section{Conclusion}
We have presented our progress towards a complete determination of $B-\bar{B}$ mixing matrix elements on DWF lattices from the JLQCD and RBC/UKQCD collaborations. Our set of ensembles allows us to control all relevant limits in a combined global fit: Two ensembles are at the physical pion mass, a pair of ensembles to study finite-volume effects and in total six different lattice spacings all the way up to $a^{-1}=4.5$ GeV. The heavy-quark masses range from below the physical charm-quark mass to about three-quarters of the bottom-quark mass. We are working on a fully non-perturbative renormalisation using the Rome-Southampton method in the RI-SMOM scheme. On each individual ensemble, a fully correlated, combined fit to two-point functions and ratios of three-point over two-point functions allows us to extract all relevant data with a single $\chi^2/$dof for each of the five mixing operators.

\section*{Acknowledgements}
The  authors  thank  the members of  the  RBC,  UKQCD and JLQCD  Collaborations  for
helpful discussions and suggestions. This work used the DiRAC Extreme Scaling service at the University of Edinburgh, operated by the Edinburgh Parallel Computing Centre on behalf of the STFC DiRAC HPC Facility (www.dirac.ac.uk). The equipment was funded by BEIS capital funding via STFC grants ST/R00238X/1 and ST/S002537/1 and STFC DiRAC Operations grantST/R001006/1. DiRAC is part of the National e-Infrastructure. F.E. and A.P. received funding from the European Research Council (ERC) under the European Union's Horizon 2020 research and innovation programme under grant agreement No 757646 \& A.P. additionally by grant agreement 813942. This research in part used computational resources provided by Multidisciplinary Cooperative Research Program in Center for Computational Sciences, University of Tsukuba and by the HPCI System Research Project (Project ID: hp210146). The work of T.K. is supported in part by JSPS KAKENHI Grant Number 21H01085. A.J. is supported by STFC grant ST/T000775/1. The project leading to this application has received funding from the European Union's Horizon 2020 research and innovation programme under the Marie Sk{\l}odowska-Curie grant agreement No 894103.
\bibliographystyle{JHEP}
\bibliography{bib}

\providecommand{\href}[2]{#2}\begingroup\raggedright\begin{thebibliography}{10}

\bibitem{Garron:2016mva}
{\scshape RBC/UKQCD} collaboration,
  \href{https://doi.org/10.1007/JHEP11(2016)001}{\emph{JHEP} {\bfseries 11}
  (2016) 001} [\href{https://arxiv.org/abs/1609.03334}{{\ttfamily
  1609.03334}}].

\bibitem{Kettle:2016mmt}
{\scshape RBC/UKQCD} collaboration,
  \href{https://doi.org/10.22323/1.256.0397}{\emph{PoS} {\bfseries LATTICE2016}
  (2016) 397} [\href{https://arxiv.org/abs/1703.00392}{{\ttfamily
  1703.00392}}].

\bibitem{Boyle:2017skn}
{\scshape RBC/UKQCD} collaboration,
  \href{https://doi.org/10.1007/JHEP10(2017)054}{\emph{JHEP} {\bfseries 10}
  (2017) 054} [\href{https://arxiv.org/abs/1708.03552}{{\ttfamily
  1708.03552}}].

\bibitem{Boyle:2017ssm}
P.~Boyle, N.~Garron, J.~Kettle, A.~Khamseh and J.~T. Tsang,
  \href{https://doi.org/10.1051/epjconf/201817513010}{\emph{EPJ Web Conf.}
  {\bfseries 175} (2018) 13010}
  [\href{https://arxiv.org/abs/1710.09176}{{\ttfamily 1710.09176}}].

\bibitem{Boyle:2018eor}
P.~Boyle, N.~Garron, R.~J. Hudspith, A.~J{\"u}ttner, J.~Kettle, A.~Khamseh
  et~al., \href{https://doi.org/10.22323/1.334.0285}{\emph{PoS} {\bfseries
  LATTICE2018} (2019) 285} [\href{https://arxiv.org/abs/1812.04981}{{\ttfamily
  1812.04981}}].

\bibitem{Albertus:2010nm}
C.~Albertus et~al.,
  \href{https://doi.org/10.1103/PhysRevD.82.014505}{\emph{Phys. Rev. D}
  {\bfseries 82} (2010) 014505}
  [\href{https://arxiv.org/abs/1001.2023}{{\ttfamily 1001.2023}}].

\bibitem{Aoki:2014nga}
Y.~Aoki, T.~Ishikawa, T.~Izubuchi, C.~Lehner and A.~Soni,
  \href{https://doi.org/10.1103/PhysRevD.91.114505}{\emph{Phys. Rev. D}
  {\bfseries 91} (2015) 114505}
  [\href{https://arxiv.org/abs/1406.6192}{{\ttfamily 1406.6192}}].

\bibitem{Boyle:2018knm}
{\scshape RBC/UKQCD} collaboration,
  \href{https://arxiv.org/abs/1812.08791}{{\ttfamily 1812.08791}}.

\bibitem{Martinelli:1994ty}
G.~Martinelli, C.~Pittori, C.~T. Sachrajda, M.~Testa and A.~Vladikas,
  \href{https://doi.org/10.1016/0550-3213(95)00126-D}{\emph{Nucl. Phys. B}
  {\bfseries 445} (1995) 81}
  [\href{https://arxiv.org/abs/hep-lat/9411010}{{\ttfamily hep-lat/9411010}}].

\bibitem{Sturm:2009kb}
C.~Sturm, Y.~Aoki, N.~H. Christ, T.~Izubuchi, C.~T.~C. Sachrajda and A.~Soni,
  \href{https://doi.org/10.1103/PhysRevD.80.014501}{\emph{Phys. Rev. D}
  {\bfseries 80} (2009) 014501}
  [\href{https://arxiv.org/abs/0901.2599}{{\ttfamily 0901.2599}}].

\bibitem{HFLAV:2019otj}
{\scshape HFLAV} collaboration,
  \href{https://doi.org/10.1140/epjc/s10052-020-8156-7}{\emph{Eur. Phys. J. C}
  {\bfseries 81} (2021) 226}
  [\href{https://arxiv.org/abs/1909.12524}{{\ttfamily 1909.12524}}].

\bibitem{ARGUS:1987xtv}
{\scshape ARGUS} collaboration,
  \href{https://doi.org/10.1016/0370-2693(87)91177-4}{\emph{Phys. Lett. B}
  {\bfseries 192} (1987) 245}.

\bibitem{CDF:2006imy}
{\scshape CDF} collaboration,
  \href{https://doi.org/10.1103/PhysRevLett.97.242003}{\emph{Phys. Rev. Lett.}
  {\bfseries 97} (2006) 242003}
  [\href{https://arxiv.org/abs/hep-ex/0609040}{{\ttfamily hep-ex/0609040}}].

\bibitem{LHCb:2011vae}
{\scshape LHCb} collaboration,
  \href{https://doi.org/10.1016/j.physletb.2012.02.031}{\emph{Phys. Lett. B}
  {\bfseries 709} (2012) 177}
  [\href{https://arxiv.org/abs/1112.4311}{{\ttfamily 1112.4311}}].

\bibitem{LHCb:2012dgy}
{\scshape LHCb} collaboration,
  \href{https://doi.org/10.1140/epjc/s10052-012-2022-1}{\emph{Eur. Phys. J. C}
  {\bfseries 72} (2012) 2022}
  [\href{https://arxiv.org/abs/1202.4979}{{\ttfamily 1202.4979}}].

\bibitem{LHCb:2012mhu}
{\scshape LHCb} collaboration,
  \href{https://doi.org/10.1016/j.physletb.2013.01.019}{\emph{Phys. Lett. B}
  {\bfseries 719} (2013) 318}
  [\href{https://arxiv.org/abs/1210.6750}{{\ttfamily 1210.6750}}].

\bibitem{LHCb:2013lrq}
{\scshape LHCb} collaboration,
  \href{https://doi.org/10.1088/1367-2630/15/5/053021}{\emph{New J. Phys.}
  {\bfseries 15} (2013) 053021}
  [\href{https://arxiv.org/abs/1304.4741}{{\ttfamily 1304.4741}}].

\bibitem{LHCb:2013fep}
{\scshape LHCb} collaboration,
  \href{https://doi.org/10.1140/epjc/s10052-013-2655-8}{\emph{Eur. Phys. J. C}
  {\bfseries 73} (2013) 2655}
  [\href{https://arxiv.org/abs/1308.1302}{{\ttfamily 1308.1302}}].

\bibitem{LHCb:2014iah}
{\scshape LHCb} collaboration,
  \href{https://doi.org/10.1103/PhysRevLett.114.041801}{\emph{Phys. Rev. Lett.}
  {\bfseries 114} (2015) 041801}
  [\href{https://arxiv.org/abs/1411.3104}{{\ttfamily 1411.3104}}].

\bibitem{Dowdall:2019bea}
R.~J. Dowdall, C.~T.~H. Davies, R.~R. Horgan, G.~P. Lepage, C.~J. Monahan,
  J.~Shigemitsu et~al.,
  \href{https://doi.org/10.1103/PhysRevD.100.094508}{\emph{Phys. Rev. D}
  {\bfseries 100} (2019) 094508}
  [\href{https://arxiv.org/abs/1907.01025}{{\ttfamily 1907.01025}}].

\bibitem{FermilabLattice:2016ipl}
{\scshape Fermilab Lattice, MILC} collaboration,
  \href{https://doi.org/10.1103/PhysRevD.93.113016}{\emph{Phys. Rev. D}
  {\bfseries 93} (2016) 113016}
  [\href{https://arxiv.org/abs/1602.03560}{{\ttfamily 1602.03560}}].

\bibitem{Grozin:2016uqy}
A.~G. Grozin, R.~Klein, T.~Mannel and A.~A. Pivovarov,
  \href{https://doi.org/10.1103/PhysRevD.94.034024}{\emph{Phys. Rev. D}
  {\bfseries 94} (2016) 034024}
  [\href{https://arxiv.org/abs/1606.06054}{{\ttfamily 1606.06054}}].

\bibitem{Grozin:2017uto}
A.~G. Grozin, T.~Mannel and A.~A. Pivovarov,
  \href{https://doi.org/10.1103/PhysRevD.96.074032}{\emph{Phys. Rev. D}
  {\bfseries 96} (2017) 074032}
  [\href{https://arxiv.org/abs/1706.05910}{{\ttfamily 1706.05910}}].

\bibitem{Grozin:2018wtg}
A.~G. Grozin, T.~Mannel and A.~A. Pivovarov,
  \href{https://doi.org/10.1103/PhysRevD.98.054020}{\emph{Phys. Rev. D}
  {\bfseries 98} (2018) 054020}
  [\href{https://arxiv.org/abs/1806.00253}{{\ttfamily 1806.00253}}].

\bibitem{Kirk:2017juj}
M.~Kirk, A.~Lenz and T.~Rauh,
  \href{https://doi.org/10.1007/JHEP12(2017)068}{\emph{JHEP} {\bfseries 12}
  (2017) 068} [\href{https://arxiv.org/abs/1711.02100}{{\ttfamily
  1711.02100}}].

\bibitem{King:2019lal}
D.~King, A.~Lenz and T.~Rauh,
  \href{https://doi.org/10.1007/JHEP05(2019)034}{\emph{JHEP} {\bfseries 05}
  (2019) 034} [\href{https://arxiv.org/abs/1904.00940}{{\ttfamily
  1904.00940}}].

\bibitem{DiLuzio:2019jyq}
L.~Di~Luzio, M.~Kirk, A.~Lenz and T.~Rauh,
  \href{https://doi.org/10.1007/JHEP12(2019)009}{\emph{JHEP} {\bfseries 12}
  (2019) 009} [\href{https://arxiv.org/abs/1909.11087}{{\ttfamily
  1909.11087}}].

\bibitem{Kaplan:1992bt}
D.~B. Kaplan, \href{https://doi.org/10.1016/0370-2693(92)91112-M}{\emph{Phys.
  Lett. B} {\bfseries 288} (1992) 342}
  [\href{https://arxiv.org/abs/hep-lat/9206013}{{\ttfamily hep-lat/9206013}}].

\bibitem{Blum:1996jf}
T.~Blum and A.~Soni, \href{https://doi.org/10.1103/PhysRevD.56.174}{\emph{Phys.
  Rev. D} {\bfseries 56} (1997) 174}
  [\href{https://arxiv.org/abs/hep-lat/9611030}{{\ttfamily hep-lat/9611030}}].

\bibitem{Blum:1997mz}
T.~Blum and A.~Soni,
  \href{https://doi.org/10.1103/PhysRevLett.79.3595}{\emph{Phys. Rev. Lett.}
  {\bfseries 79} (1997) 3595}
  [\href{https://arxiv.org/abs/hep-lat/9706023}{{\ttfamily hep-lat/9706023}}].

\bibitem{Allton:2008pn}
{\scshape RBC/UKQCD} collaboration,
  \href{https://doi.org/10.1103/PhysRevD.78.114509}{\emph{Phys. Rev.}
  {\bfseries D78} (2008) 114509}
  [\href{https://arxiv.org/abs/0804.0473}{{\ttfamily 0804.0473}}].

\bibitem{PhysRevD.93.074505}
{\scshape RBC/UKQCD} collaboration,
  \href{https://doi.org/10.1103/PhysRevD.93.074505}{\emph{Phys. Rev. D}
  {\bfseries 93} (2016) 074505}.

\bibitem{Boyle:2017jwu}
P.~A. Boyle, L.~Del~Debbio, A.~J{\"u}ttner, A.~Khamseh, F.~Sanfilippo and J.~T.
  Tsang, \href{https://doi.org/10.1007/JHEP12(2017)008}{\emph{JHEP} {\bfseries
  12} (2017) 008} [\href{https://arxiv.org/abs/1701.02644}{{\ttfamily
  1701.02644}}].

\bibitem{Nakayama:2016atf}
K.~Nakayama, B.~Fahy and S.~Hashimoto,
  \href{https://doi.org/10.1103/PhysRevD.94.054507}{\emph{Phys. Rev. D}
  {\bfseries 94} (2016) 054507}
  [\href{https://arxiv.org/abs/1606.01002}{{\ttfamily 1606.01002}}].

\bibitem{Morningstar:2003gk}
C.~Morningstar and M.~J. Peardon,
  \href{https://doi.org/10.1103/PhysRevD.69.054501}{\emph{Phys. Rev. D}
  {\bfseries 69} (2004) 054501}
  [\href{https://arxiv.org/abs/hep-lat/0311018}{{\ttfamily hep-lat/0311018}}].

\bibitem{Brower:2012vk}
R.~C. Brower, H.~Neff and K.~Orginos,
  \href{https://doi.org/10.1016/j.cpc.2017.01.024}{\emph{Comput. Phys. Commun.}
  {\bfseries 220} (2017) 1} [\href{https://arxiv.org/abs/1206.5214}{{\ttfamily
  1206.5214}}].

\bibitem{Boyle:2016lbp}
P.~A. Boyle, G.~Cossu, A.~Yamaguchi and A.~Portelli,
  \href{https://doi.org/10.22323/1.251.0023}{\emph{PoS} {\bfseries LATTICE2015}
  (2016) 023, \url{https://github.com/paboyle/Grid}}.

\bibitem{Portelli:Hadrons}
A.~Portelli, {\emph{\url{https://github.com/aportelli/Hadrons}} }.

\bibitem{GUSKEN1989266}
S.~Güsken, U.~Löw, K.-H. Mütter, R.~Sommer, A.~Patel and K.~Schilling,
  \href{https://doi.org/https://doi.org/10.1016/S0370-2693(89)80034-6}{\emph{Physics
  Letters B} {\bfseries 227} (1989) 266}.

\bibitem{ALEXANDROU199160}
C.~Alexandrou, F.~Jegerlehner, S.~Güsken, K.~Schilling and R.~Sommer,
  \href{https://doi.org/https://doi.org/10.1016/0370-2693(91)90219-G}{\emph{Physics
  Letters B} {\bfseries 256} (1991) 60}.

\bibitem{UKQCD:1993gym}
{\scshape UKQCD} collaboration,
  \href{https://doi.org/10.1103/PhysRevD.47.5128}{\emph{Phys. Rev. D}
  {\bfseries 47} (1993) 5128}
  [\href{https://arxiv.org/abs/hep-lat/9303009}{{\ttfamily hep-lat/9303009}}].

\end{thebibliography}\endgroup

\end{document}